\documentclass[%
 reprint,
 amsmath,amssymb,
 aps,
]{revtex4-2}

\usepackage{amsmath}
\usepackage{graphicx}
\usepackage{dcolumn}
\usepackage{bm}

\usepackage[utf8]{inputenc}
\usepackage[T1]{fontenc}
\usepackage{mathptmx}
\usepackage[T1]{fontenc} 
\usepackage{amsmath,color}
\usepackage{amssymb}
\usepackage{amsfonts}
\usepackage{amsthm}
\usepackage{mathtools}
\usepackage[title]{appendix}
\usepackage{hyperref}
\usepackage{braket}
\usepackage{xcolor}
\usepackage{etoolbox}
\usepackage{soul}
\usepackage{mathrsfs}
\usepackage[mathcal]{eucal}
\usepackage{tabularx,booktabs,array}
\newcolumntype{Y}{>{\centering\arraybackslash}X}

\newcommand\footnoteref[1]{\protected@xdef\@thefnmark{\ref{#1}}\@footnotemark}
\makeatother

\definecolor{change}{RGB}{0,0,255}

\usepackage{algorithm}
\usepackage{algpseudocode}
\usepackage{dsfont}

\usepackage{bbold}

\renewcommand\thefigure{\arabic{figure}}
\begin{document}
\preprint{AIP/123-QED}

\title{Effective reorganization energy for electron transfer}
\author{Ethan Abraham}
\email{ethana@mit.edu}
\affiliation{Department of Chemistry, Massachusetts Institute of Technology, Cambridge, Massachusetts 02139, USA}
\author{Junghyun Yoon}
\affiliation{Department of Chemical Engineering, Massachusetts Institute of Technology, Cambridge, Massachusetts 02139, USA}
\author{Troy Van Voorhis}
\affiliation{\mbox{Department of Chemistry, Massachusetts Institute of Technology, Cambridge, Massachusetts 02139, USA}}
\author{Martin Z. Bazant}
\affiliation{Department of Chemical Engineering, Massachusetts Institute of Technology, Cambridge, Massachusetts 02139, USA}
\affiliation{Department of Mathematics, Massachusetts Institute of Technology, Cambridge, Massachusetts 02139, USA}

\date{\today}

\begin{abstract}
The Marcus theory expression for the rate of non-adiabatic electron transfer is widely used across a range of physical conditions. Although Marcus theory defines the reorganization energy classically, here we show that the reorganization parameter appearing in the activation barrier for normal-region electron transfer is most generally a quantum mechanical object that depends on the electronic coupling, coinciding with the Marcus picture only in the limit of vanishing electronic coupling. This result unifies the physical description of electron-transfer activation barriers across the adiabatic and non-adiabatic regimes and formally predicts that Marcus-like rate expressions remain accurate beyond their traditional non-adiabatic domain of validity. These insights allow us to derive a closed-form expression for the curvature of the current--overpotential relation for electron-transfer-limited reactions at the electrochemical interface, now formally applicable to both inner-sphere and outer-sphere processes.
\end{abstract}

\maketitle

The fundamental physical quantity in the theory of condensed phase electron transfer (ET) has been the reorganization energy. Defined by Marcus as the energy required to bring the equilibrium nuclear coordinates of one charge state to those of the other in the absence of ET, the operational meaning of the reorganization energy has remained essentially unchanged for decades \cite{Marcus_1956,Marcus_1985,Nobel_Marcus,Nitzan_2006,Ratner_review,Bredas_Review,Bazant_CIET}.

However, a recent and previously unrecognized discrepancy has emerged in studies of the carbon dioxide reduction reaction (CO$_2$RR): the reorganization energy inferred from experimental kinetics \cite{zhang_driving_2020,Lees2024} is nearly an order of magnitude smaller than that predicted theoretically by \textit{ab initio} simulations within Marcus theory \cite{Qin2023_CO2_Reduction,ethan_CIET,experimental_footnote}. This discrepancy persists despite the fact that the reaction is observed experimentally to follow Marcus-like rate expressions with remarkable accuracy \cite{zhang_driving_2020,Lees_CO2R_crossover,king_revealing_2025,Brown2020ChemRxiv}. These observations suggest that the reorganization energy extracted from experiment may be an ontologically distinct quantity from the classical reorganization energy defined in Marcus theory. Here, we resolve this discrepancy and show that it arises from the fundamental structure of the harmonic two-level quantum system.

We recall that Marcus theory in it's most basic formulation assumes that the reorganization energy---and by extension the activation barrier---can be determined classically, with quantum mechanics entering only through the prefactor via a hopping rate \cite{Marcus_1956,Marcus_1985,Nitzan_2006}. While extensions beyond these simple assumptions have been widely explored \cite{prefactor_Gladkikh,Jianshu_spectral,friction_Garg,rates_Lawrence,Fay_recent,Nitzan_2006}, in this Letter we demonstrate a surprising mathematical property of the harmonic two-level quantum system, which implies that in the normal region the reorganization parameter appearing in the activation barrier depends not only on the classical Marcus reorganization energy but also on the electronic coupling. In addition to explaining the CO$_2$RR discrepancy and persistence of Marcus-like rate expressions---where the electronic coupling is thought to be strong \cite{ethan_CIET,Qin2023_CO2_Reduction,Koper2024Theory}---this result challenges the classical picture of ET activation and implies that the reorganization energy inferred from experiment is a coupling-dependent quantity. We show that within the Condon approximation \cite{Marcus_1985,Blumberger_Condon,Nitzan_2006}, the \textit{effective reorganization energy} ($\lambda_\text{eff}$) appearing in the Marcus-like expression for the activation barrier is given by
\begin{equation}\label{simplest}
\lambda_{\mathrm{eff}}=\lambda\left(1-\frac{2V}{\lambda}\right)^2,
\end{equation}
which recovers the classical reorganization energy 
($\lambda$) in the limit of vanishing electronic coupling ($V$), but extends the formal range of validity of Marcus-like rate expressions beyond the traditional non-adiabatic limit to strongly adiabatic reactions \cite{Abraham2025AdiabaticET}. Our results provide a unified quantum description of electron-transfer activation barriers for both inner-sphere and outer-sphere reactions across a broad range of electronic coupling strengths. Importantly, the results that follow imply that the reorganization energies fitted from experimental kinetics should be interpreted as $\lambda_\text{eff}$ rather than $\lambda.$

{\it Effective reorganization energy.} We begin by recalling a rudimentary mathematical treatment of the two-level harmonic quantum system \cite{Ratner_review,Marcus_1985,Brunschwig_Book,Bredas_Review,Fay_recent,Nitzan_2006} described by two diabatic electronic states $\ket{\phi_a}$ and $\ket{\phi_b}$ whose energies are given by \begin{subequations}\label{energies}\begin{equation}\label{Ea} E_a(q)=\lambda q^2,\end{equation}\begin{equation}\label{Eb}E_b(q)=\lambda(1-q)^2+\Delta E.\end{equation}\end{subequations} Here, $\lambda $ is the diabatic reorganization energy defined in Marcus theory (classical, unperturbed by the electronic coupling), $q$ is a collective nuclear coordinate defined to satisfy $\langle q\rangle_a=0, \langle q\rangle_b=1$, and $\Delta E$ is the driving force. Setting the two energies equal, we obtain the coordinate and energy of the transition state (TS,*) in the non-adiabatic (i.e. $V\rightarrow0$) limit, \begin{equation}\label{marcus_ts}q^*=\frac{1}{2}\left(1+\frac{\Delta E}{\lambda}\right),\end{equation}\begin{equation}\label{marcus_Ea}E^*(\Delta E)=\frac{(\lambda+\Delta E)^2}{4\lambda},\end{equation} a central result of Marcus theory \cite{Marcus_1956,BixonJortner1999,Bredas_Review,Ratner_review,BixonJortner1999,Nitzan_2006}.
  
\begin{figure}[!hbt]
\includegraphics[width=1\columnwidth]{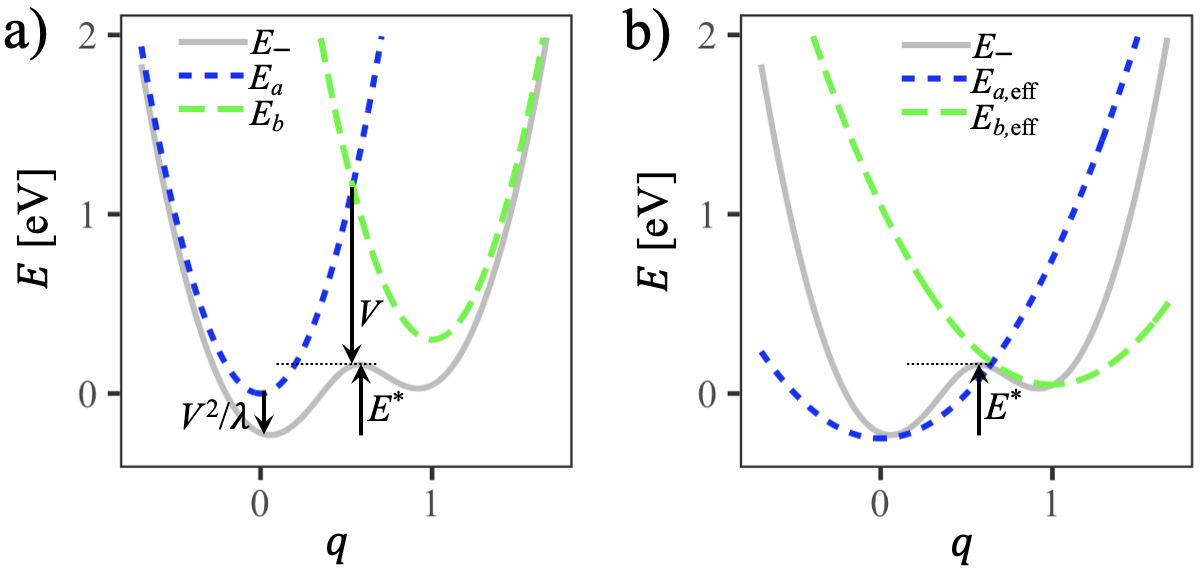}
\caption{(a) Energy $E$ plotted against the reaction coordinate $q$ for the two-level harmonic ET system with parameter values taken to be $\lambda=4.0$ eV, $\Delta E=0.3$ eV. Marcus diabats with harmonic energies $E_a$ (blue dashed) and $E_b$ (green dashed) are mixed by large constant coupling $V=1.0$ eV, giving the ground state adiabat $E_-$ (grey solid). (b) Same as (a) except now showing the auxiliary parabolas $E_{a,\text{aux}}$ (blue dashed) and $E_{b,\text{aux}}$ (green dashed) whose curvatures are given by $\lambda_\text{eff}=1.0$ eV and whose intersection leads mathematically to the correct activation barrier. (c) Reaction pathways through two-dimensional space of nuclear coordinate $q$ and electronic mixing. Non-adiabatic limit (zig-zag dotted line): sequential motion where nuclei move first, then ET occurs, then nuclei relax. Adiabatic regime (gray solid curve): concerted nuclear-electronic motion along smooth path.}
\label{FIG1}
\end{figure} 

In general, the diabatic states will mix via an electronic coupling $V(q)$ \cite{Nitzan_2006,Blumberger_Condon,Jianshu_spectral,Fay_recent,VanVoorhis_2010,Newton_probes,Hush_1997,Hush_1999}. We can therefore also cast Eq. (\ref{energies}) as the two-level Hamiltonian in the diabatic basis $\{\ket{\phi_a},\ket{\phi_b}\}$\begin{equation} \label{matrix}H(q)=
\begin{pmatrix}
E_a(q) & V(q) \\
V(q) & E_b(q)
\end{pmatrix}=\begin{pmatrix}
\lambda q^2 & V(q) \\
V(q) & \lambda(1-q)^2+\Delta E
\end{pmatrix},
\end{equation} which upon solving the secular equation gives the energies of the corresponding states in the adiabatic basis $\{\ket{\psi_-
},\ket{\psi_+}\}$ \begin{equation}\label{adiabats}\begin{split}E_\pm(q)&=\frac{E_a(q)+E_b(q)}{2}\pm\frac{1}{2}\sqrt{(E_a(q)-E_b(q))^2+4V(q)^2}\\&=\frac{\lambda(2q^2-2q+1)+\Delta E}{2}\pm\\&~~~~~~~~~~~~~~~~~~~~~~~~~\frac{1}{2}\sqrt{(\lambda(2q-1)-\Delta E)^2+4V(q)^2}
.\end{split}\end{equation} The exact activation barrier in this model is given by \begin{equation}\label{Ea_exact}E^*=E_-(q^*)-E_-(q_\text{r}),\end{equation} where $q^*$ is the coordinate of the TS and $q_\text{r}\approx0$ the coordinate of the reactant minimum.

It is noteworthy that up until this point, we have only reproduced a standard treatment of a well-established model. At the same time, it is well known phenomenologically that the activation barrier satisfies \begin{equation}\label{result} E^*(\Delta E)=\frac{(\lambda_{\rm eff}+\Delta E)^2}{4\lambda_{\rm eff}}\end{equation} to a good approximation, especially for $-\lambda_\text{eff}<\Delta E<\lambda_\text{eff}$ \cite{experiment_Weller,experimental_Silverstein,experimental_Foote,experiment_Scheerer,experimental_Beitz,experimental_Sun,Ratner_review,Nobel_Marcus,Nitzan_2006}. Here, $\lambda_\text{eff}$ is a parameter which we call the \textit{effective reorganization energy}. The first fundamental result of this Letter is that---within the minimal model of Eq. (\ref{energies})--(\ref{result})---there has been a subtle but consequential category error in the identification of $\lambda_\text{eff}$ that appears in Eq. (\ref{result}) with the $\lambda$ of Eq. (\ref{energies}). Such an identification is perturbatively inconsistent, for it retains terms to the quadratic order in $\Delta E$ but not in $V$ \cite{griffiths_schroeter_2018}. The quadratic expansion of Eq. (\ref{Ea_exact}) is \cite{Blumberger_expression,Brunschwig_Book} \begin{equation}\label{ts}\begin{split} E^*=\frac{\lambda}{4}+\frac{\Delta E}{2}&+\frac{(\Delta E)^2}{4\lambda}-V(q^*)+\frac{V(q_\text{r})^2}{\lambda} +\\&O\left( \frac{V_\text{max}^3}{\lambda^2}, \frac{V_\text{max}^2 \Delta E}{\lambda^2}, \frac{V_\text{max} (\Delta E)^2}{\lambda^2}, \frac{(\Delta E)^3}{\lambda^2} \right), \end{split}\end{equation} where $V_\text{max}=\text{max}_{q \in [0,1]}V(q)$ is defined to collect higher order terms (see Supplementary Material for further derivation). While the first three terms retained are familiar from Eq. (\ref{marcus_Ea}), the final two terms enter at the same perturbative order \cite{Hush_adiabatic,Nitzan_2006,Zener1932,vonNeumannWigner1929,Newton_probes,Bazant_CIET,ethan_CIET,Brunschwig_Book,griffiths_schroeter_2018} and therefore should also be retained. 

The second fundamental result is the existence of a unique (to quadratic order) $\lambda_\text{eff}$ such that Eq. (\ref{result}) correctly expands to Eq. (\ref{ts}). It is given by \begin{equation}\label{eff} \lambda_{\rm eff}=\lambda-4V(q^*)+\frac{4V(q_\text{r})^2}{\lambda}. \end{equation} If one invokes the Condon approximation that $V(q)=V$, which was commonly employed by Marcus \cite{Marcus_1985,Marcus_1989,Nitzan_2006,Blumberger_Condon,Troy_Condon}, the situation is further simplified, and Eq. (\ref{eff}) reduces to Eq. (\ref{simplest}). 

It is crucial to note that although Eqs. (\ref{simplest}) and ({\ref{eff}}) are introduced here as an ansatz to make the commonly used Eq. (\ref{result}) consistent with second order perturbation theory, it is shown in Ref. \cite{Abraham2025AdiabaticET} that the term $(\Delta E)^2/4\lambda_\text{eff}$ according to which Eq. (\ref{result}) differs from the truncated component of Eq. (\ref{ts}) constitutes a summation of higher order terms making the former significantly more accurate than the latter.

Although the present treatment differs from the well-established theory by only the previous three paragraphs, profound consequences emerge. First, the assumption given in Ref. \cite{Marcus_1956} that vanishing electronic coupling is a necessary condition for the emergence of Eq. (\ref{result}) does not hold. Secondly, the paradox of the inconsistent reorganization energies in the studies of CO$_2$RR is immediately resolved. For corresponding simulation and experiment, Qin \textit{et al.} reported $\lambda = 6.3$~eV from \textit{ab initio} calculations \cite{Qin2023_CO2_Reduction}, whereas
Zhang \textit{et al.} extracted $\lambda_{\text{eff}} = 0.75$~eV from experimental current--overpotential
data that is indeed well-described by Eq. (\ref{result}) \cite{zhang_driving_2020,experimental_footnote}.
 Inverting Eq. (\ref{simplest}) now gives \begin{equation}\label{approxV}V=\frac{\lambda}{2}-\frac{1}{2}\sqrt{\lambda^2-\lambda(\lambda-\lambda_\text{eff})}.\end{equation} Using the reported values in Eq. (\ref{approxV}), we obtain $V\approx2$ eV, which is consistent with our own \textit{ab initio} calculations of the coupling value \cite{ethan_CIET}. Finally, as established in Ref. \cite{Abraham2025AdiabaticET}, the normal region is defined approximately by $-\lambda_\text{eff}<\Delta E$ whereas the region $-\lambda<\Delta E<-\lambda_\text{eff}$ is barrierless. As we discuss in that work \cite{Abraham2025AdiabaticET}, this might help explain why Marcus's predictions for the inverted region (i.e. $\Delta E <-\lambda$) \cite{Nobel_Marcus,Marcus_1985} are observed in some systems \cite{experiment_Miller,experimental_Giovanny,experimental_Sun,experiment_Scheerer}, but not others \cite{experiment_Weller,experimental_Beitz}
 
In the example of CO$_2$RR, the fact that the current–overpotential relationship for a complex inner-sphere reaction can be described by Eq. (\ref{matrix})---a two-level matrix model yielding a quantitative solution in terms of only three parameters $\lambda$, $V$, and $\Delta E$---is remarkable, and suggests that a physically minimal formalism will be more robust in describing the activation of complex and strongly-coupled systems than originally anticipated---at least once the effective reorganization energy is appropriately defined.

Figure 1 illustrates how the present picture differs from the standard Marcus construction for non-negligible electronic coupling. The intersection of the Marcus curves shown in Fig. 1(a) severely overestimates the true activation barrier, which is instead obtained by intersecting the auxiliary parabolas shown in Fig. 1(b) with curvatures set by the effective reorganization energy $\lambda_\text{eff}$. We clarify that these auxiliary parabolas are mathematical constructs whose intersection leads to the correct calculation of the activation barrier as $\Delta E$ varies; therefore, they do not have the same interpretation as the diabatic free energy curves in Marcus theory. At the same time, our mathematical result implies that distinct dynamics can underlie the same phenomenological rate expression. As emphasized in Fig. 1(c), the reaction may be viewed as occurring in a two-dimensional space of the classical nuclear coordinate and quantum electronic mixing. Whereas the non-adiabatic Marcus theory assumes a zig-zag path through this space in which nuclei and electrons move sequentially, strong coupling enables concerted nuclear–electronic motion along a smooth trajectory, reducing the effective reorganization energy. In both of these dynamical extremes, the dependence on the driving force is described by Eq. (\ref{result}). 

\begin{figure}[!hbt]
\includegraphics[width=1\columnwidth]{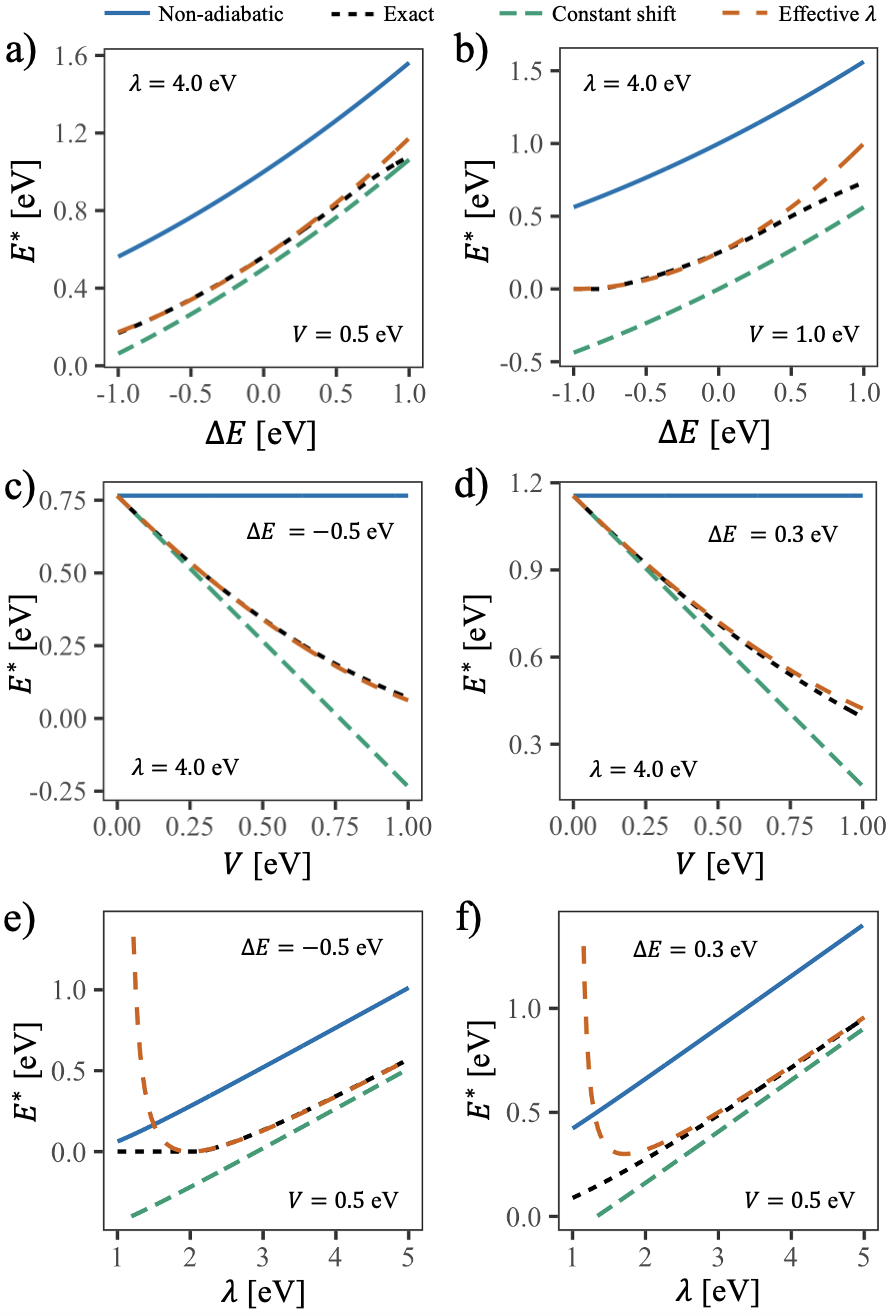}
\caption{Comparison of activation barriers $E^*$ predicted by non-adiabatic Marcus theory (blue solid), the exact ground-state adiabat (black dotted), the traditional constant-shift correction (green dashed), and the effective reorganization energy introduced here (orange dashed). Panels show the dependence of $E^*$ on the driving force $\Delta E$, coupling $V$, and diabatic reorganization energy $\lambda$.}
\label{FIG2}
\end{figure}

Next, we demonstrate the quantitative accuracy of our interpretation for systems described by Eq. (\ref{matrix}). In Fig. 2, we compare the exact activation barrier, obtained from numerical evaluation of Eqs.~(\ref{adiabats})--(\ref{Ea_exact}), with several common approximations across a range of $\lambda$, $V$, and $\Delta E$. (For clarity, we adopt the Condon approximation $V(q)=V$ in these examples; the Supplementary Materials show that the approach generalizes robustly for several non-constant couplings $V(q)$ \cite{Lockwood_nonCondon,Skourtis_nonCondon,Troy_Condon,Ratner_Condon}.) In all six panels, the standard non-adiabatic Marcus barrier substantially overestimates the true barrier at large coupling. A constant $-V$ correction, accounting only for the lowering of the diabatic crossing \cite{Hush_adiabatic,Nitzan_2006,Zener1932,vonNeumannWigner1929,Newton_probes,Bazant_CIET}, improves the agreement but exhibits a significant underestimation of the barrier when $V$ is large. In contrast, Eqs. (\ref{result})-(\ref{eff}) reproduce the exact barrier with excellent agreement across a broad range of parameter values. 

These results show that for systems adequately described by the present model, Eqs.~(\ref{result})--(\ref{eff}) are quantitatively accurate over most of the experimentally accessible normal region. The approximation becomes exact when either $\Delta E=0$ or $V=0$, but remains accurate over a much broader parameter range. For sufficiently large $|\Delta E|$, higher-order terms in Eq.~(\ref{ts}) become significant and the approximation ultimately breaks down. However, as shown in Fig.~2(a)--(b), excellent agreement persists for substantial positive $\Delta E$, and especially for $\Delta E<0$. In practice, the approximation is quantitatively accurate over the full range $-\lambda_{\mathrm{eff}} \le \Delta E \le 0$. For the coupling parameter $V$, a singular limit is approached as $V \rightarrow \lambda/2$, where $\lambda_{\mathrm{eff}} \rightarrow 0$. In this limit,
\begin{equation}
E^* \xrightarrow[]{V\rightarrow\frac{1}{2}\lambda} 0
,\end{equation}
consistent with the known condition $V \ge \lambda/2$ for barrierless electron transfer \cite{Jianshu_spectral,Hush_1997,Hush_1999,Brunschwig_Book}. Within this broad domain of validity, Eqs. (\ref{result})--(\ref{eff}) provide a closed-form interpolation between the non-adiabatic and strongly adiabatic regimes, a result that remains especially accurate after integration over the band structure of a metal.

While this Letter is motivated by electrochemistry, we expect that clarifying the difference between $\lambda$ and $\lambda_\text{eff}$ will be important for charge-transfer systems in other fields as well. For example, in studies of disordered organic semiconductors \cite{Troy_OLED}, mixed valence complexes \cite{Zhu2021LandauZenerCrossover}, and hole transfer in DNA $\pi$-stacks \cite{2009DNAChargeTransfer}, coupling values have been reported within an order of magnitude of the reorganization energy. This suggests that the effect of the electronic coupling on the effective reorganization energy discussed here is likely to be a general and unavoidable property of ET across a variety of systems described by coupled harmonic states \cite{Marcus_Holstein_PRB,Plaron_Hopping,Troy_OLED,Marcus_Organic_Semiconductor,Marcus_biology,2009DNAChargeTransfer}.

{\it ET-limited Faradaic reactions.} We now show how the above framework can be used to derive a rate expression for ET-limited Faradaic reactions, which has a similar mathematical structure to Marcus--Hush--Chidsey (MHC) kinetics \cite{Chidsey,MHC_Simple,Bazant_CIET}, but is valid well beyond the non-adiabatic limit and applies to both inner- and outer-sphere reactions.

The theory of heterogeneous ET at the electrochemical interface treats ET from a continuum of states varying in energy $\epsilon$ (defined relative to the Fermi level in the electrode) to a discrete state (i.e. an orbital of a molecule in the electrolyte) \cite{Gurney_1931}. Traditionally, the rate of the reduction reaction is obtained from integrating over the continuum of contributing rates \begin{equation}\label{genmhc}\begin{aligned}
k_\text{red/ox}=A\int d\epsilon \rho(\epsilon)n(\epsilon)\exp\left[-\beta E^*(\pm\Delta E(\epsilon))\right],
\end{aligned}\end{equation} where $A$ is a prefactor, $\rho$ is the density of states in the continuum, $n$ is the Fermi-Dirac occupancy, $\beta=1/k_BT$ is the inverse thermal energy, and $E^*(\Delta E)$ is the activation energy of Eq. (\ref{marcus_Ea}), which depends on the integration variable $\epsilon$ through \begin{equation}\label{epsover}\begin{aligned}\Delta E(\epsilon)=e\eta_\text{f}-\epsilon.
\end{aligned}\end{equation} Here $e$ is the elementary charge, and $\eta_\text{f}$ is the formal overpotential (defined so a negative value favors reduction of the atom/molecule by the electrode) \cite{Bazant_CIET}. 
\begin{figure}[!hbt]
\includegraphics[width=1\columnwidth]{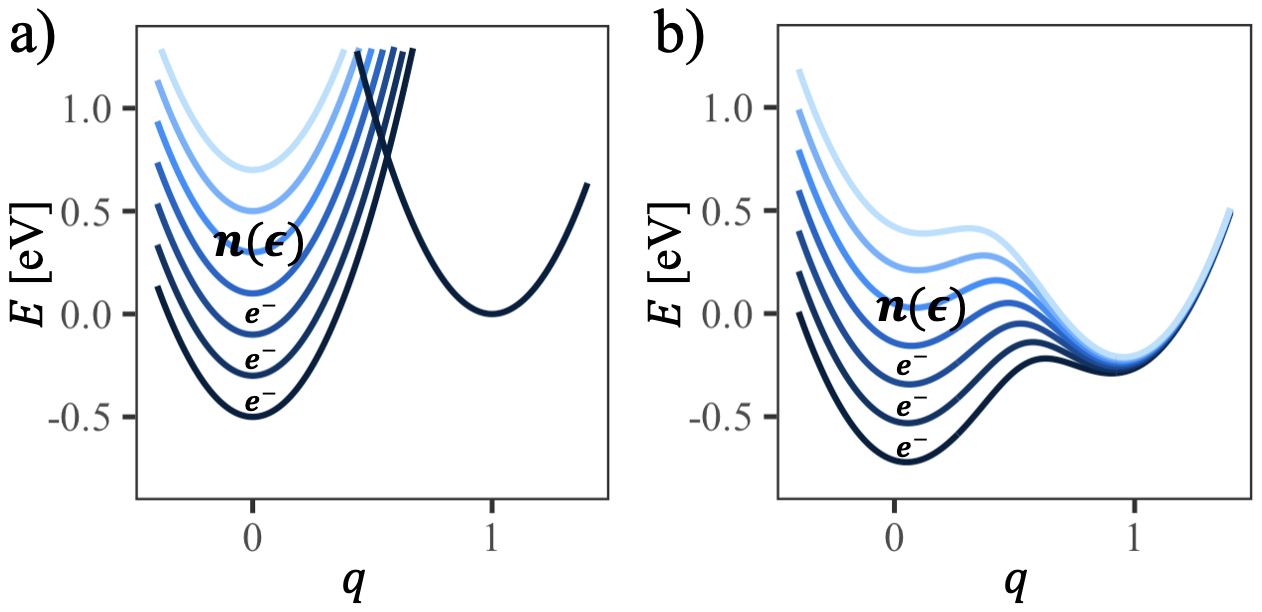}
\caption{Schematic comparison of heterogeneous ET from a metal to a discrete state in (a) the non-adiabatic MHC picture and (b) the adiabatic picture introduced here. In both cases, the total rate is a sum of rates between many donor states with Fermi-Dirac occupancy $n(\epsilon)$ and a final acceptor state, but the barrier for each transition is changed significantly due to the coupling. This is reflected mathematically by taking $\lambda\rightarrow\lambda_\text{eff}$ in Eq. (\ref{genmhc}).}
\label{FIG3}
\end{figure}

 \begin{figure}[!hbt]
\includegraphics[width=1\columnwidth]{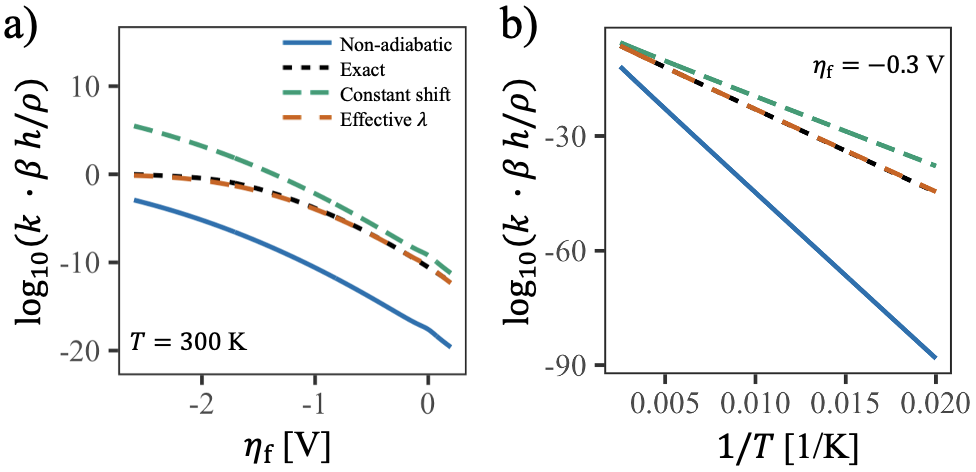}
\caption{Comparison of the reduction currents $k=k_\text{red}$ predicted by non-adiabatic Marcus theory (blue solid), the exact ground-state adiabat (black dotted), the traditional constant-shift correction (green dashed), and the effective reorganization energy introduced here (orange dashed). (a) Logarithm of the (scaled) rate plotted against the formal overpotential $\eta_\text{f}$ at $T=300$ K. (b) Arrhenius plot showing the dependence of the rate on the inverse temperature $1/T$ for fixed formal overpotential $\eta_\text{f}=-0.3$ V. In both panels the reorganization energy and coupling are set to $\lambda=4.0$ eV and $V=0.5$ eV respectively.}
\label{FIG4}
\end{figure}

We argue that the necessity of integrating over the continuum of possible transitions \cite{Gurney_1931}, though originally applied for non-adiabatic ET \cite{Marcus_1965,Chidsey,MHC_Simple}, should also apply for the adiabatic case, as suggested by Fig. 3. The difference is that whereas in the non-adiabatic limit the prefactor $A$ is given by $A^\text{non-adiabatic}=(|V|^2/\hbar)\sqrt{\pi\beta/\lambda}$ derived from the Fermi-Golden rule \cite{Nitzan_2006,Marcus_1985}, in the adiabatic limit it will be replaced by the classical attempt frequency $A^\text{adiabatic}=\nu\approx1/\beta h.$ A closed-form expression of Eq. (\ref{genmhc}) for non-adiabatic Marcus theory has been derived in Ref. \cite{MHC_Simple}, using the wide-band approximation $\rho(\epsilon)\approx\rho$, and evaluating the energy integral that combines the Gaussian activation factor with the Fermi–Dirac occupancy. The corresponding expression can now be derived for the adiabatic limit by substituting $A^\text{non-adiabatic}\rightarrow A^\text{adiabatic}$ and $\lambda\rightarrow\lambda_\text{eff},$ whereupon we obtain
\begin{equation}\label{mhcapprox}
k_\text{red/ox}=\frac{\rho\sqrt{\pi\lambda_\text{eff}/\beta}}{\beta h\left(1 + e^{\pm\beta e \eta_\text{f}}\right)}
 \mathrm{erfc}\left(
 \frac{\beta\lambda_\text{eff}- \sqrt{1 + \sqrt{\beta\lambda_\text{eff}} + (\beta e \eta_\text{f})^2}}
      {2\sqrt{\beta \lambda_\text{eff}}}
 \right).
 \end{equation} Note that in contrast to the corresponding non-adiabatic form of this equation \cite{MHC_Simple,Bazant_CIET}, the coupling no longer appears in the prefactor, and $\sqrt{\lambda_\text{eff}}$ appears in the numerator of the prefactor because there is no cancellation by the normalized Franck-Condon factor. Note that in general, an ion-transfer (IT) barrier must also be considered \cite{fraggedakis_theory_2021,Bazant_CIET,ethan_CIET,CIET_LCO}; here we focus on the ET component of the rate.

We can now compare the behavior of this rate equation to the standard non-adiabatic rate expression \cite{MHC_Simple}, as well as to the traditional adiabatic correction via a constant $-V$ barrier lowering, and finally to the exact rate using Eqs. (\ref{adiabats})--(\ref{Ea_exact}) in Eq. (\ref{genmhc}) and evaluating the integral numerically. Figure 4(a) shows the logarithm of scaled rate against the formal overpotential, for an example system at room temperature. Figure 4(b) fixes the formal overpotential and shows an Arrhenius plot for the same system. Remarkably, we observe excellent agreement between the rate calculated with the effective reorganization energy and the numerically calculated exact rate over all experimentally relevant temperatures and overpotentials. In fact, the deviations observed in Fig. 2(a)-(b) at high $\Delta E$ are observed not to affect the overall rate due to the negligible contribution of such endothermic transitions, while any deviations at very large negative $\Delta E$ (i.e. the inverted region) are suppressed by the Fermi-Dirac factor $n(\epsilon)$. Both non-adiabatic Marcus theory and the constant $-V$ shift correction perform much worse than the correction proposed here. As shown in the Supplementary Material, this analysis can be applied outside the Condon approximation with a similar level of success, simply using Eq. (\ref{eff}) instead of Eq. (\ref{simplest}).

In conclusion, we have demonstrated that the reorganization parameter entering the Marcus-like expression for the activation barrier is a quantum mechanical quantity that depends on the electronic coupling and coincides with the Marcus picture only in the non-adiabatic limit. The use of Marcus-like rate expressions for adiabatic electrochemical reactions is now formally justified, which explains why Marcus-like ECIT rate expressions \cite{Bazant_CIET} remain accurate for Faradaic reactions that often involve strong electronic coupling \cite{CIET_LCO,CIET_LFP,zhang_driving_2020,Bazant_CIET}. Furthermore, our results help clarify the physical origin of Tafel’s law \cite{Tafel_original,Gurney_1931,Butler_1936}, which arises when IT rather than ET dominates the activation barrier \cite{Bazant_CIET}. Specifically, strong electronic coupling reduces the effective reorganization energy, causing the IT component to control the barrier. Finally, because both the reorganization energy and the electronic coupling can be computed \textit{ab initio} \cite{ethan_CIET,Troy_mixed_cdft,Blumberger_Condon}, Eqs. (\ref{eff}) and (\ref{mhcapprox}) provide a formal basis for first-principles prediction of the ET contribution to Faradaic reaction kinetics \cite{ethan_CIET}. 

\section*{\NoCaseChange{Acknowledgments}} E.A. is grateful to Abraham Nitzan and Daniel Nocera for useful discussions regarding the formal correctness and experimental relevance of the proposed theory. This work was supported by the MIT Faculty Research Innovation Fund (FRIF) through the generosity of Irv and Melinda Simon.

\section*{\NoCaseChange{Data Availability}} Sample code used to generate the above numerical analysis of the two-level quantum harmonic system is available at \url{https://github.com/eabes23/effective_lambda}.

\makeatletter
\setlength{\parindent}{15pt}
\@afterindenttrue
\makeatother
\widetext
\begin{center}
\textbf{\large Supplementary Material:\\Effective reorganization energy for electron transfer}
\end{center}
\setcounter{equation}{0}
\setcounter{figure}{0}
\setcounter{table}{0}
\setcounter{section}{0}
\setcounter{page}{1}
\makeatletter
\renewcommand{\theequation}{S\arabic{equation}}
\renewcommand{\thefigure}{S\arabic{figure}}
\renewcommand{\bibnumfmt}[1]{[S#1]}
\renewcommand{\citenumfont}[1]{S#1}
\renewcommand{\thepage}{S\arabic{page}} 

\section{Equation (\ref{ts}) as an application of perturbation theory} Although one can roughly convince oneself of the main result (i.e. Eqs. (\ref{result})--(\ref{eff}) of the main text) in several lines of algebra, a full proof requires an extensive calculation because the reaction coordinate of the reactant minimum and transition state will in general shift slightly due to the coupling. This proof will be covered in the following section; here we show that under the simplifying assumption that $q_\text{r}=0$ and $q^*=q^*_\text{non-adiabatic}$ with constant ($q$-independent) coupling, Eq. (\ref{ts}) can be understood using standard perturbation theory, keeping the first non-vanishing correction to both the reactant basin and transition state. 

Define $H=H_0+H'$ with \begin{align}\label{matrix_pert1}&H_0(q)=\begin{pmatrix}
\lambda q^2 & 0 \\
0 & \lambda(1-q)^2+\Delta E
\end{pmatrix}
,\\&H'(q)=\begin{pmatrix}
0 & V \\
V & 0
\end{pmatrix},\end{align} where we have used the Condon approximation $V(q)=V.$ The first order correction at $q=0$ is \begin{align}\label{matrix_pert2_0}&E^{(1)}_0(0)=\bra{\psi_a(q=0)}H'\ket{\psi_a(q=0)}=\begin{pmatrix}1&0\end{pmatrix}\begin{pmatrix}
0 & V \\
V & 0
\end{pmatrix}\begin{pmatrix}1\\0\end{pmatrix}=0
.\end{align} Taking the second order correction, \begin{equation}\label{second_order}E^{(2)}_0(0)=\frac{|\bra{\psi_b(q=0)}H'\ket{\psi_a(q=0)}|^2}{E_a(0)-E_b(0)},\end{equation} we obtain
\begin{equation}\begin{split}\label{second_order_continued}E^{(2)}_0(0)&=-\frac{V^2}{\lambda+\Delta E}\\&=-\frac{V^2}{\lambda}+O\left(\frac{V^2\Delta E}{\lambda^2}\right).\end{split}\end{equation} 

Now for $q=q^*,$ we have the degeneracy $E_a(q^*)=E_b(q^*)=E^*_\text{non-adiabatic}$, and so to obtain the first order correction we must use the eigenstates of $H'$. We obtain \begin{align}\label{matrix_pert2_TST}&E^{(1)}_0(q^*)=\bra{\psi_-(q=q^*)}H'\ket{\psi_-(q=q^*)}=\frac{1}{\sqrt{2}}\begin{pmatrix}1&-1\end{pmatrix}\begin{pmatrix}
0 & V \\
V & 0
\end{pmatrix}\frac{1}{\sqrt{2}}\begin{pmatrix}1\\-1\end{pmatrix}=-V
.\end{align} Thus we have\begin{equation}\begin{split}\label{pert_result}E^*&=[E^{(0)}_0(q^*)+E_0^{(1)}(q^*)]-[E^{(0)}_0(0)+E_0^{(2)}(0)]\\&=\left[\frac{\lambda}{4}+\frac{\Delta E}{2}+\frac{(\Delta E)^2}{4\lambda}-V\right]-\left[0-\frac{V^2}{\lambda} +O\left(\frac{V^2\Delta E}{\lambda^2}\right)\right],\end{split}\end{equation} which agrees with Eq. (\ref{ts}) as desired.

\section{Derivation of Equations (\ref{ts})-(\ref{eff}) for constant coupling}

Here we derive Eq. (\ref{ts})--(\ref{eff}) in the Condon approximation $V(q)=V$, and in the following section we show that the derivation extends to a more general $V(q)$. The exact activation energy is given by Eq. (\ref{Ea_exact}). The non-adiabatic limit of Marcus theory is obtained by taking $V\rightarrow0$ in Eq. (\ref{matrix}); in this case, $E_-(q)=\text{min}\{E_a(q),E_b(q)\}$ where $E_a(q)$ and $E_b(q)$ are defined in Eq. (\ref{energies}). We obtain the standard result of non-adiabatic Marcus theory \begin{equation}\label{diabatic_crossing}q^*_{\text{non-adiabatic}}=\frac{1}{2}\left(1+\frac{\Delta E}{\lambda}\right),\end{equation} and \begin{equation}\label{standard}E^*_{\text{non-adiabatic}}=\frac{(\lambda+\Delta E)^2}{4\lambda},\end{equation} where we have used $q_\text{r}=0$, which is exact in this limit.

Now consider the case when $V>0.$ It is an exact result from diagonalization of the Eq. (\ref{matrix}) that $E_-(q^*_{\text{non-adiabatic}})=E^*_{\text{non-adiabatic}}-V$. Combining this relation with the assumption that the location of the transition state and reaction state is unchanged by the coupling, and also the assumption that the energy of the reactant basin is not lowered by the coupling, leads to the standard "constant shift" adiabatic correction to Marcus theory \begin{equation}\label{lowering}E^*_{\text{constant-shift}}=\frac{(\lambda+\Delta E)^2}{4\lambda}-V.\end{equation}

Corrections to Eq. (\ref{lowering}) arise when (i) the stationary points of $E_-(q)$ shift due to the coupling (i.e. $q_\text{r}\neq 0,~q^*\neq q^*$) and (ii) the coupling $V$ is large enough that the lowering of the reactant basin (i.e. $E_-(q_\text{r})<0$) becomes significant. We show that (ii) is addressed by the correction proposed in this Letter, while (i) gives rise to higher order corrections that can be neglected when $\Delta E,V\ll\lambda,$ as is often the case in the Marcus normal regime.

The strategy to prove Eq. (\ref{ts})--(\ref{eff}) is to introduce small shifts $\delta_\text{r}$, $\delta^*$ to account for any deviation in the location of the reactant basin and transition state in the adiabatic landscape as compared to the corresponding basins of the diabatic landscape. We compute the transition barrier according to Eq.~(\ref{Ea_exact}), keeping terms consistently up to $O\left(V^2/\lambda,\; (\Delta E)^2/\lambda,\; V\Delta E/\lambda^2\right)$ and discarding
$O\left(V^k(\Delta E)^{3-k}/\lambda^2\right)$ for $k = 0,1,2,3$. This is valid when $\lambda$ is relatively large compared to $\Delta E$ and $V$.

\subsection{Local expansion around diabatic crossing point}
We first determine the energy of the adiabatic transition state by performing a local expansion around the diabatic crossing point, $q_\text{c}$. It is convenient to introduce the sum and difference of the diabatic energies, $\Sigma(q)=E_a(q)+E_b(q)$ and $\Delta(q)=E_a(q)-E_b(q)$. Then the adiabatic eigenvalues from Eq.~(\ref{adiabats}) can be written compactly as
\begin{equation}\label{compact}
    E_\pm(q) = \tfrac{1}{2}\Sigma(q) \pm \tfrac{1}{2}\sqrt{\Delta(q)^2+4V^2}
\end{equation}
The diabatic crossing point $q_\text{c}$ is defined by $\Delta(q_\text{c})=0$ where $q_\text{c} = q_\text{non-adiabatic}^*$. The adiabatic transition state $q^*$ is located near $q_\text{c}$, so we express it as a small displacement, $q^*=q_\text{c}+\delta^*$. Since $\Delta(q)=\lambda(2q-1)-\Delta E$ is linear, the shift is exact
\begin{equation}
\Delta(q^*)=2\lambda\,\delta^*.
\end{equation}
For the quadratic sum \(\Sigma(q)=\lambda(2q^2-2q+1)+\Delta E\), a direct
shift gives the exact identity
\begin{equation}\label{qstar}
\Sigma(q^*)=\Sigma(q_\text{c})
+2\Delta E\,\delta^*+2\lambda(\delta^*)^2,
\end{equation}
where
\begin{equation}\label{qc}
\Sigma(q_\text{c})=\frac{\lambda}{2}+\Delta E
+\frac{(\Delta E)^2}{2\lambda}.
\end{equation}

Inserting Eq. (\ref{qstar}) with Eq. (\ref{qc}) into the ground state solution of Eq. (\ref{compact}), we obtain
\begin{equation}\label{eq:Em_cross_local}
    E_-(q^*) = \tfrac{1}{2}\Sigma(q_\text{c}) + \Delta E \delta^* + \lambda (\delta^*)^2 - \frac{\sqrt{4\lambda^2 (\delta^*)^2 + 4 V^2}}{2}.
\end{equation} Expanding the square root for small $\delta^*$,
\begin{equation}
V\sqrt{1+\left(\frac{\lambda \delta^*}{V}\right)^2}=V+\frac{\lambda^2 (\delta^*)^2}{2V}
+O\left(\frac{\lambda^4 (\delta^*)^4}{V^3}\right)
\end{equation}
gives
\begin{equation}\label{eq:Em_cross_local_bigO}
    E_-(q^*)= \tfrac{1}{2}\Sigma(q_\text{c})-V+ \Delta E \delta^* + \left(\lambda-\frac{\lambda^2}{2V}\right)(\delta^*)^2 + O\left(\frac{\lambda^4(\delta^*)^4}{V^3}\right).   
\end{equation} Stationarity $dE_-/d\delta^*=0$ now yields
\begin{equation}\label{eq:deltaTS}
\delta^*_{\rm TS} = -\frac{\Delta E}{2\lambda-\lambda^2/V}.
\end{equation}
Substituting Eq. \eqref{eq:deltaTS} back into \eqref{eq:Em_cross_local_bigO}, the extremum value is
\begin{equation}\label{S19}
    E_-(q^*) = \tfrac{1}{2}\Sigma(q_\text{c})-V
- \frac{(\Delta E)^2}{4\lambda-2\lambda^2/V}.
\end{equation} For bookkeeping in powers of $V/\lambda$ and $\Delta E/\lambda$,
\begin{equation}\label{S20}
\frac{1}{4\lambda-2\lambda^2/V}
=-\frac{V}{2\lambda^2}\frac{1}{1-2V/\lambda}
=\frac{-V}{2\lambda^2}\left[1+O\left(\frac{V}{\lambda}\right)\right].
\end{equation}
Therefore,
\begin{equation}\label{eq:ETS_scaling}
E_-(q^*)=\frac{\lambda}{4}+\frac{\Delta E}{2}+\frac{(\Delta E)^2}{4\lambda}
- V + O\left(\tfrac{V(\Delta E)^2}{\lambda^2},\tfrac{V(\Delta E)^4}{\lambda^4}\right).
\end{equation}

\subsection{Local expansion around reaction minimum}
We proceed analogously to the above by writing $q_\text{r} = \delta$ with $\delta$ small near the reactant well. Using the same $\Sigma$ and $\Delta$ notation, we have
\begin{equation}
\begin{split}
    \Sigma(q_\text{r}) &= \lambda(2\delta^2-2\delta+1)+\Delta E, \\
    \Delta(q_\text{r}) &= \lambda(2\delta-1)-\Delta E.
\end{split}
\end{equation} Thus,
\begin{equation} \label{eq:Er_local_exact}
        E_-(q_\text{r}) = \frac{\lambda+\Delta E}{2} - \lambda \delta +\lambda \delta^2 - \frac{\sqrt{(-(\lambda+\Delta E)+2\lambda\delta)^2+4V^2}}{2}
\end{equation}
Evaluating Eq.~\eqref{eq:Er_local_exact} at $\delta=0$ gives 
\begin{equation} \label{eq:Er0_exact}
    E_-(0) = \frac{\lambda+\Delta E}{2} -\frac{\sqrt{(\lambda+\Delta E)^2+4V^2}}{2}.
\end{equation}
Expanding the square root for $V\ll\lambda+\Delta E$,
\begin{equation}\label{eq:E0_root_expand}
        \frac{\lambda+\Delta E}{2}\sqrt{1+\left(\frac{2V}{\lambda+\Delta E}\right)^2}=
        \frac{\lambda+\Delta E}{2}+\frac{V^2}{\lambda+\Delta E} +O\left(\frac{V^4}{\lambda^3}\right),
\end{equation}
so the adiabatic mixing lowers the energy at $q=0$ by
\begin{equation}\label{eq:Er0_series}
E_-(0)=-\frac{V^2}{\lambda+\Delta E}
+O\left(\frac{V^4}{\lambda^3}\right).
\end{equation}
To exhibit the Marcus normal regime ($\lambda\gg\Delta E$) explicitly, we expand the prefactor
\begin{equation}\label{S27}
\frac{1}{\lambda+\Delta E}
=\frac{1}{\lambda}\left(1+\frac{\Delta E}{\lambda}\right)^{-1}
=\frac{1}{\lambda}-\frac{\Delta E}{\lambda^2}
+O\left(\frac{(\Delta E)^2}{\lambda^3}\right)
\end{equation}
which gives the desired asymptotic form
\begin{equation}\label{eq:Er0_series_lambda_normal}
E_-(0)=-\frac{V^2}{\lambda}
+O\left(\frac{V^2\,\Delta E}{\lambda^2},\frac{V^2(\Delta E)^2}{\lambda^3},\frac{V^4}{\lambda^3}\right).
\end{equation} 

Next we show that the shift of the minimum is negligible at our order of truncation. To accomplish this, we perform a Taylor expansion of the square-root term in Eq.~\eqref{eq:Er_local_exact} in powers of $\delta$ about $\delta=0$. Defining $S_0 = \tfrac{1}{2} \sqrt{\left(\lambda+\Delta E\right)^2+4V^2}$, we find that
\begin{equation}
    \sqrt{\left(-\frac{\lambda+\Delta E}{2}+\lambda\delta\right)^2+V^2}= S_0 - \frac{\lambda(\lambda+\Delta E)}{2S_0}\delta
    + \left(\frac{\lambda^2 V^2}{2S_0^3}\right)\delta^2 +O(\delta^3).
\end{equation}
Inserting this into Eq.~\eqref{eq:Er_local_exact} and collecting the constant, linear, and quadratic terms in $\delta$
\begin{equation}\label{eq:Er_quad_form}
E_-(\delta)=E_-(0)+L\delta+Q\delta^2+O(\delta^3),
\end{equation}
we find that
\begin{equation} \label{eq:Er_quad}
    L=-\lambda + \frac{\lambda(\lambda+\Delta E)}{2S_0}
    = -\lambda+\frac{\lambda(\lambda+\Delta E)}{\sqrt{(\lambda+\Delta E)^2+4V^2}},
\end{equation}
\begin{equation}
    Q = \lambda -\frac{\lambda^2 V^2}{2S_0^3}
    =\lambda
    -\frac{4\lambda^2V^2}{\left((\lambda+\Delta E)^2+4V^2\right)^{3/2}}.
\end{equation}
Now we  expand $L$ and $Q$ for $V \ll \lambda+\Delta E$ as we did in Eq.~\eqref{eq:E0_root_expand}. Using the expansions \begin{equation}
\frac{\lambda(\lambda+\Delta E)}{\sqrt{(\lambda+\Delta E)^2+4V^2}}
= \lambda\left[1-\frac{2V^2}{(\lambda+\Delta E)^2}\right]
+ O\!\left(\frac{V^4}{\lambda^3}\right),
\end{equation}
and
\begin{equation}
\frac{4\lambda^2V^2}{\big((\lambda+\Delta E)^2+4V^2\big)^{3/2}}
= \frac{4\lambda^2V^2}{(\lambda+\Delta E)^3}\left[1-\frac{6V^2}{(\lambda+\Delta E)^2}\right]
+ O\!\left(\frac{V^4}{\lambda^3}\right),
\end{equation}
we obtain
\begin{equation}
    L=-\frac{2\lambda V^2}{(\lambda+\Delta E)^2}+O\left(\frac{V^4}{\lambda^3}\right)=O\left(\frac{V^2}{\lambda}\right)
\end{equation} and
\begin{equation}
Q = \lambda
- \frac{4\lambda^2V^2}{(\lambda+\Delta E)^3}
+ O\!\left(\frac{V^4}{\lambda^3}\right)=\lambda+O\left(\frac{V^2}{\lambda}\right).
\end{equation}

Finally, recall that the reactant minimum satisfies $dE_-/d\delta=0$, so from Eq.~\eqref{eq:Er_quad_form} we obtain
\begin{equation}\label{eq:delta_r_value}
    \delta_{\text{r}}=-\frac{L}{2Q}
    =\frac{V^2}{(\lambda+\Delta E)^2}\,\frac{\lambda}{Q}
    =O\left(\frac{V^2}{\lambda^2}\right).
\end{equation}
Its energetic effect is
\begin{equation}\label{eq:delta_r_energy}
    E_-(\delta_{\text{r}})-E_-(0)=-\frac{L^2}{4Q}
    =-\frac{\lambda^2 V^4}{Q(\lambda+\Delta E)^4}
    =O\left(\frac{V^4}{\lambda^3}\right),
\end{equation}
which is beyond the retained orders. Therefore, to the accuracy of
Eq.~\eqref{eq:Er0_series_lambda_normal} we can set
\begin{equation}\label{eq:Er_min_final}
    E_-(q_\text{r})=E_-(0)=-\frac{V^2}{\lambda}
    +O\left(\frac{V^2\Delta E}{\lambda^2},
    \frac{V^2(\Delta E)^2}{\lambda^3},
    \frac{V^4}{\lambda^3}\right).
\end{equation}
Using Eq. (\ref{eq:Er_min_final}) and Eq. (\ref{eq:ETS_scaling}) in Eq. (\ref{Ea_exact}) gives Eq. (\ref{ts}), as desired.
\\
\subsection{Agreement of $\lambda_\text{eff}$ approximation with exact barrier}

Having proven Eq. (\ref{ts}), it remains to be shows that using Eq. (\ref{eff}) in Eq. (\ref{result}) gives an asymptotically equivalent result. Equation (\ref{result}) is equivalent to \begin{equation}\begin{split}\label{eq:approx_barrier}
E^*&=\frac{\lambda_\text{eff}}{4}+\frac{\Delta E}{2}+\frac{(\Delta E)^2}{4\lambda_\text{eff}}\\&=\left(\frac{\lambda}{4}-V+\frac{V^2}{\lambda}\right)+\frac{\Delta E}{2}+\frac{(\Delta E)^2}{4\left(\frac{\lambda}{4}-V+\frac{V^2}{\lambda}\right)}.\end{split}
\end{equation}
Comparing Eq. (\ref{eq:approx_barrier}) with Eq. (\ref{ts}), we need only to show that \begin{equation}\label{eq:final_step1}
\frac{(\Delta E)^2}{4\lambda}-\frac{(\Delta E)^2}{4\left(\lambda-4V+\frac{4V^2}{\lambda}\right)}=O\left(
\frac{V^3}{\lambda^2},
\frac{V^2 \Delta E}{\lambda^2},
\frac{V (\Delta E)^2}{\lambda^2},
\frac{(\Delta E)^3}{\lambda^2}
\right).\end{equation}
To prove Eq. (\ref{eq:final_step1}), we can Taylor expand the second term of the left hand side in $V$, which gives \begin{equation}\label{eq:expand_diff1}
\frac{(\Delta E)^2}{4\left(\lambda-4V+\frac{4V^2}{\lambda}\right)}=\frac{(\Delta E)^2}{4\lambda}+\frac{(\Delta E)^2V}{\lambda^2}+O\left(\frac{(\Delta E)^2V^2}{\lambda^3},\frac{(\Delta E)^2V^3}{\lambda^4}\right).
\end{equation} This completes the proof.

Importantly, Eq. (\ref{eq:expand_diff1}) can be used to easily show uniqueness of $\lambda_\text{eff}$ (to quadratic order) after using Eq. (\ref{eff}) in Eq. (\ref{result}) and matching coefficients with Eq. (\ref{ts}).

\section{Beyond the Condon Approximation: $\lambda_\text{eff}$ for Systems with Variable Coupling}

In practice, one often finds that the Condon approximation does not hold, that is $V(q)\neq V$ for all $q$. We claim that if the function of $V(q)$ is reasonably behaved (such as in Eq. (\ref{linV})), the results of this Letter still hold, now with \begin{equation}\label{eq:eff_nonCondon}
\lambda_{\text{eff}}=\lambda-4V(q^*)+\frac{4V(q_\text{r})^2}{\lambda}.
\end{equation}
It is clear that \begin{equation}\label{ts_no_Condon}\begin{split}
E^*=\frac{\lambda}{4}+\frac{\Delta E}{2}&+\frac{(\Delta E)^2}{4\lambda}-V(q^*)+\frac{V(q_\text{r})^2}{\lambda} 
+\\&O\left(
\frac{V_\text{max}^3}{\lambda^2},
\frac{V_\text{max}^2 \Delta E}{\lambda^2},
\frac{V_\text{max} (\Delta E)^2}{\lambda^2},
\frac{(\Delta E)^3}{\lambda^2}
\right)
\end{split}\end{equation}
where $V_\text{max}=\text{max}_{q \in [0,1]}V(q)$, so we must only show that Eq. (\ref{result}), using Eq. (\ref{eff}), agrees with Eq. (\ref{ts_no_Condon}) up to the same order of truncation. Here Eq. (\ref{result}) gives \begin{equation}\begin{split}\label{eq:approx_barrier2}
E^*&=\frac{\lambda_\text{eff}}{4}+\frac{\Delta E}{2}+\frac{(\Delta E)^2}{4\lambda_\text{eff}}\\&=\left(\frac{\lambda}{4}-V(q^*)+\frac{V(q_\text{r})^2}{\lambda}\right)+\frac{\Delta E}{2}+\frac{(\Delta E)^2}{4\left(\frac{\lambda}{4}-V(q^*)+\frac{V(q_\text{r})^2}{\lambda}\right)}.\end{split}
\end{equation}
Comparing Eq. (\ref{eq:approx_barrier2}) with Eq. (\ref{ts_no_Condon}), we need only to show that \begin{equation}\begin{split}\label{eq:final_step2}
\frac{(\Delta E)^2}{4\lambda}-&\frac{(\Delta E)^2}{4\left(\lambda-4V(q^*)+\frac{4V(q_\text{r})^2}{\lambda}\right)}=\\&O\left(
\frac{V_\text{max}^3}{\lambda^2},
\frac{V_\text{max}^2 \Delta E}{\lambda^2},
\frac{V_\text{max} (\Delta E)^2}{\lambda^2},
\frac{(\Delta E)^3}{\lambda^2}
\right).\end{split}\end{equation}
To prove Eq. (\ref{eq:final_step2}), we can Taylor expand the second term of the left hand side in small $V(q^*),V(q_\text{r})$, which gives \begin{equation}\begin{split}\label{eq:expand_diff2}
\frac{(\Delta E)^2}{4\left(\lambda-4V(q^*)+\frac{4V(q_\text{r})^2}{\lambda}\right)}=\frac{(\Delta E)^2}{4\lambda}+&\frac{(\Delta E)^2V(q^*)}{\lambda^2}-\frac{4(\Delta E)^2V(q_\text{r})^2}{\lambda^3}+\\&O\left(\frac{(\Delta E)^2V_\text{max}^2}{\lambda^3},\frac{(\Delta E)^2V_\text{max}^3}{\lambda^4}\right).
\end{split}\end{equation} This completes the proof.

In practice, often $V(0)$ and $V(1)$ will be known (e.g. from ab initio calculations), and $V(q^*)$ can be approximated by assuming $V(q)$ takes a particular functional form. In this case,  $\lambda_\text{eff}$ can be obtained by setting $V(q_\text{r})\approx V(0)$ and \begin{equation}\label{V_q_approx}V(q^*)\approx V(q^*_\text{non-adiabatic})=V(0)+\frac{1}{2}\left(1+\frac{\Delta E}{\lambda}\right)[V(1)-V(0)].\end{equation} 

In Fig. S1, we perform the analysis shown in Fig. 2 of the main text, only now for several choices of $V(q)$ (i.e. linear and quadratic) given by
\begin{equation}
\label{linV}V(q)=V(0)+q[V(1)-V(0)],
\end{equation}
\begin{equation}
\label{quadV}V(q)=V(0)+q^2[V(1)-V(0)].
\end{equation} The top panel (Fig. S1(a)-(b)) shows results for the linear choice of $V(q)$, while the bottom panel (Fig. S1(c)-(d)) shows the corresponding results for the quadratic choice. For both cases, the left panels show the activation barriers computed with all four methods described in the main text when the couplings $V(0)$ and $V(1)$ are fixed to $0.6$ eV and $1.0$ eV respectively, while the thermodynamic driving force $\Delta E$ is varied. The right panels show results when $\Delta E$ is fixed to $-0.5$ eV, $V(1)$ is fixed to $1.0$ eV, and  $V(0)$ is varied. The excellent agreement between the barrier calculated with the effective reorganization energy and the true barrier calculated from the ground state adiabat suggests that Eq. (8)-(9) from the main text are similarly valid outside the Condon approximation.

\begin{figure}[!hbt]
\includegraphics[width=0.6\columnwidth]{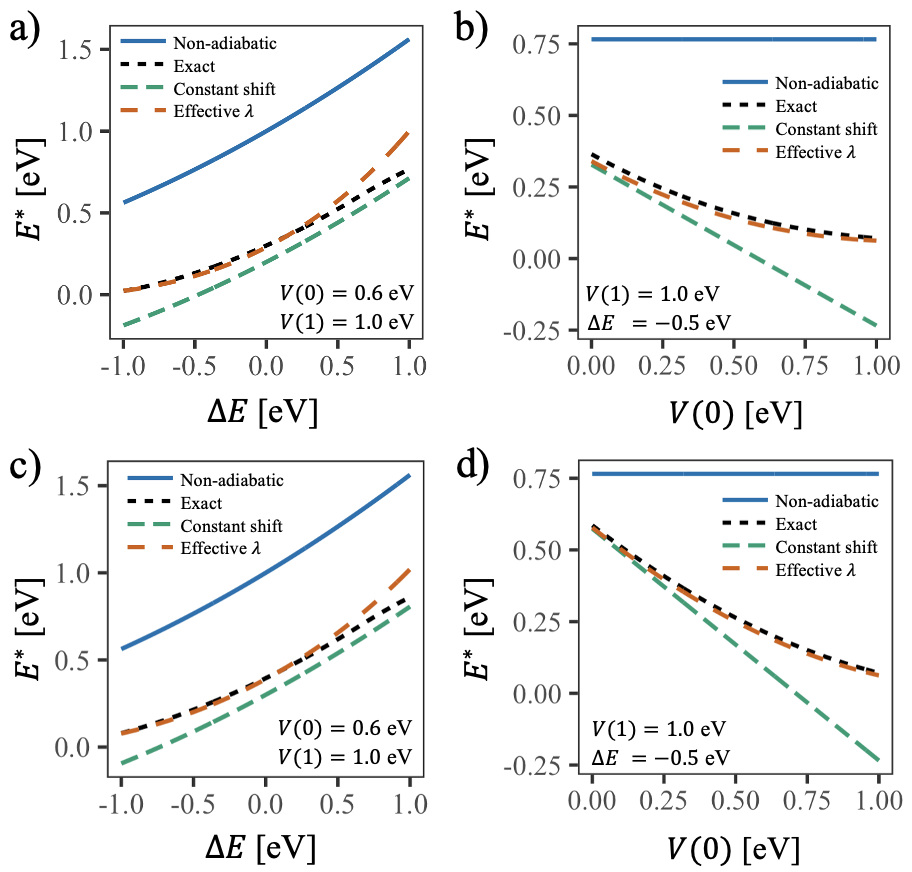}
\caption{(a) Activation barrier $E^*$ plotted against the thermodynamic driving force $\Delta E$ for $\lambda=4.0$ eV and $V(0)=0.6$ eV and $V(1)=1.0$ eV using linear $V(q).$ (b) Activation barrier $E^*$ plotted against the coupling value $V(0)$ for $V(1)=1$ eV and $\Delta E=-0.5$ eV using linear $V(q)$. (c)-(d) Same as (a)-(b), except using quadratic $V(q)$. In all panels results are shown for non-adiabatic Marcus theory with no mixing (blue solid line), the exact ground state adiabat (black dotted line), the traditional adiabatic correction to Marcus theory obtained by lowering the barrier by $V(q^*)$ (green dashes) and the correction considered in this work (orange dashes).}
\label{FIGS1}
\end{figure}

\section{Kinetics for Systems with Variable Coupling}

For the general case where the Condon approximation does not hold, if we approximate that $q_\text{r}=0$ and use $q^*$ from Eq. (\ref{marcus_ts}) in Eq. (\ref{eff}), we obtain
\begin{equation}\label{explicit_eff}\lambda_\text{eff}=\lambda-4V\left(\frac{\lambda+\Delta E}{2\lambda}\right)+\frac{4V(0)^2}{\lambda},\end{equation} which can be evaluated explicitly if the functional form $V(q)$ is known.

For the case of ET from a metal continuum, using Eq. (\ref{epsover}) from the main text in Eq. (\ref{explicit_eff}) gives
\begin{equation}\label{explicit_eff_eta}\lambda_\text{eff}(\eta_\text{f},\epsilon)\approx\lambda-4V\left(\tfrac{\lambda+e\eta_\text{f}-\epsilon}{2\lambda}\right)+\frac{4V(0)^2}{\lambda},\end{equation} which interestingly implies that $\lambda_\text{eff}$ is a function of both the formal overpotential and the variable of integration $\epsilon.$ The full rate expression, generalizing Eq. (\ref{genmhc}) for the case of variable coupling (and taking the only the reduction current for simplicity), is then given by \begin{equation}\label{genmhc2}\begin{aligned}
k=&A^\text{adiabatic}\int d\epsilon \rho(\epsilon)n(\epsilon)\exp\left[-\beta\frac{\left(\lambda_\text{eff}+\Delta E \right)^2}{4\lambda_\text{eff}} \right]\\&=\frac{k_BT}{h}\int d\epsilon \rho(\epsilon)n(\epsilon)\exp\left[-\beta\frac{\left(\lambda-4V\left(\frac{\lambda+e\eta_\text{f}-\epsilon}{2\lambda}\right)+\frac{4V(0)^2}{\lambda}+e\eta_\text{f}-\epsilon \right)^2}{4\left(\lambda-4V\left(\frac{\lambda+e\eta_\text{f}-\epsilon}{2\lambda}\right)+\frac{4V(0)^2}{\lambda}\right)} \right],
\end{aligned}\end{equation} where we have used Eq. (\ref{explicit_eff_eta}) and Eq. (\ref{epsover}) in Eq. (\ref{genmhc}). We note that the additional dependencies on $\epsilon$ through $\lambda_\text{eff}$ precludes exact application of the method presented in Ref. \cite{MHC_Simple}, where the analytical form Eq. (\ref{mhcapprox}) is not achieved exactly. However, we find that accurate results are obtained by neglecting the $\epsilon$--dependence of $\lambda_\text{eff}$ by setting $\epsilon=0,$ which is not unreasonable due to the suppression of transitions far from the Fermi level, either due to endothermicity or lack of occupancy. That is, we use \begin{equation}\label{explicit_eff_eta}\lambda_\text{eff}(\eta_\text{f})\approx\lambda-4V\left(\tfrac{\lambda+e\eta_\text{f}}{2\lambda}\right)+\frac{4V(0)^2}{\lambda}.\end{equation} 

 \begin{figure}[!hbt]
\includegraphics[width=0.6\columnwidth]{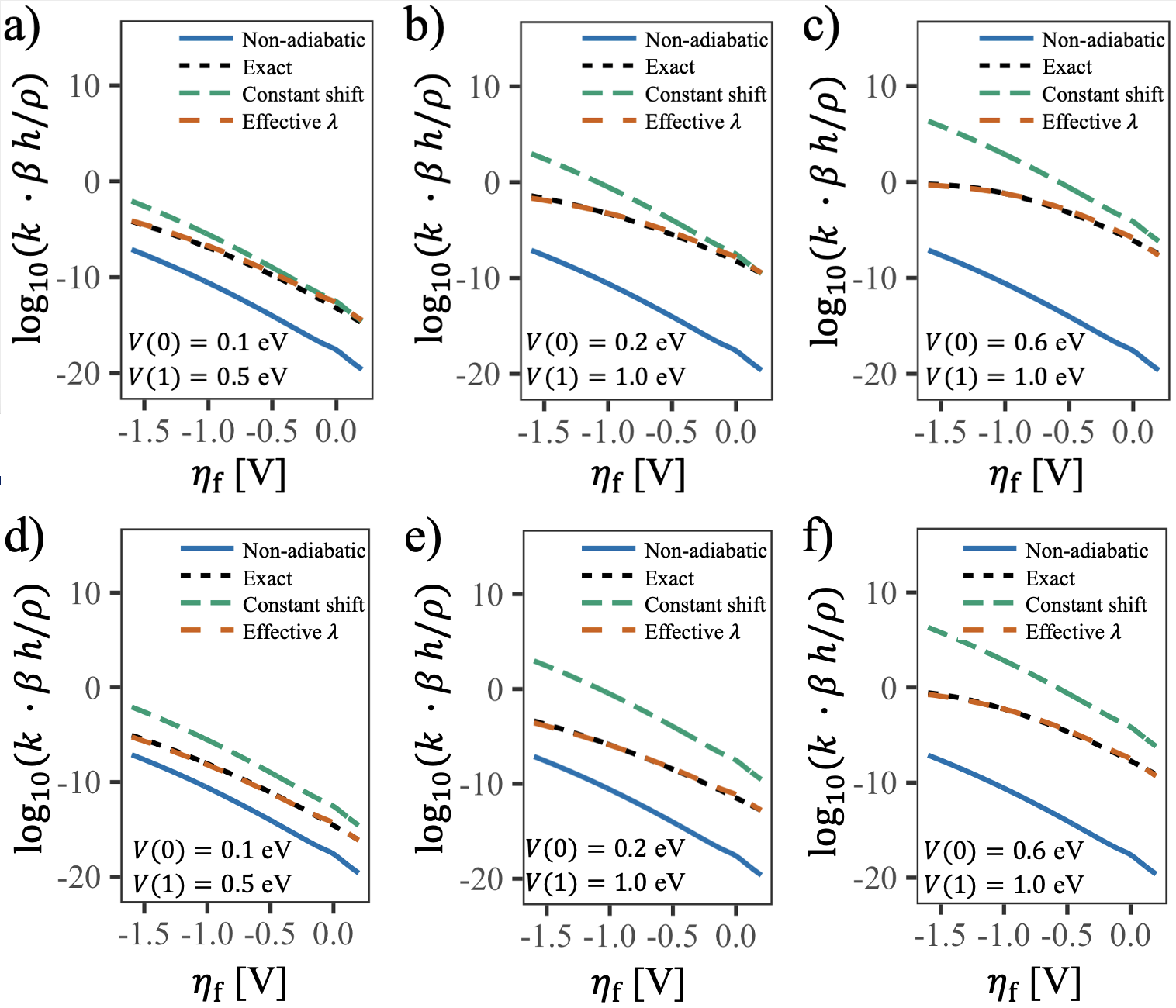}
\caption{Tafel plots showing the (base-10) logarithm of the (scaled) reaction rate $k$ plotted against the formal overpotential $\eta_\text{f}$ at $T=300$ K. Results are shown for (a)--(c) the linear $V(q)$ given by Eq. (\ref{linV}) and for (d)--(f) quadratic $V(q)$ given by Eq. (\ref{quadV}). Across each row of panels, the value $V(0)$ increased while the value of $V(1)$ is held fixed (note that plots (c) and (f) are equivalent). The orange curve is computed using Eq. (\ref{mhcapprox2})}
\label{FIGS2}
\end{figure}

Using this definition of the reorganization parameter in Eq. (\ref{mhcapprox}), we obtain \begin{equation}\label{mhcapprox2}
 k=\frac{\rho\sqrt{\frac{\pi}{\beta}\left(\lambda-4V\!\left(\frac{\lambda+e\eta_\text{f}}{2\lambda}\right)
+\frac{4V(0)^2}{\lambda}\right)}}{\beta h\left(1 + e^{\beta e \eta_\text{f}}\right)}
 \mathrm{erfc}\!\left(
\frac{\beta\left(\lambda-4V\!\left(\frac{\lambda+e\eta_\text{f}}{2\lambda}\right)
+\frac{4V(0)^2}{\lambda}\right)
- \sqrt{1 + \sqrt{\beta\left(\lambda-4V\!\left(\frac{\lambda+e\eta_\text{f}}{2\lambda}\right)
+\frac{4V(0)^2}{\lambda}\right)} + (\beta e \eta_\text{f})^2}}
      {2\sqrt{\beta \left(\lambda-4V\!\left(\frac{\lambda+e\eta_\text{f}}{2\lambda}\right)
+\frac{4V(0)^2}{\lambda}\right)}}
 \right).
\end{equation}

 The resulting Tafel plots for this case are shown in Fig. S2, where similar accuracy to the Condon approximation in the main text is observed. We compare the results using (Fig. S2(a)--(d)) the linear $V(q)$ from Eq. (\ref{linV}) and (Fig. S2(e)--(f)) the quadratic $V(q)$ from Eq. (\ref{quadV}). The panels from left to right show results for decreasing difference between $V(0)$ and $V(1)$. In all cases, an exceptional agreement is obtained between the orange curve, determined by Eq. (\ref{mhcapprox2}), and the black curve which represents the true theoretical Tafel plot. 

  \begin{figure}[!hbt]
\includegraphics[width=0.6\columnwidth]{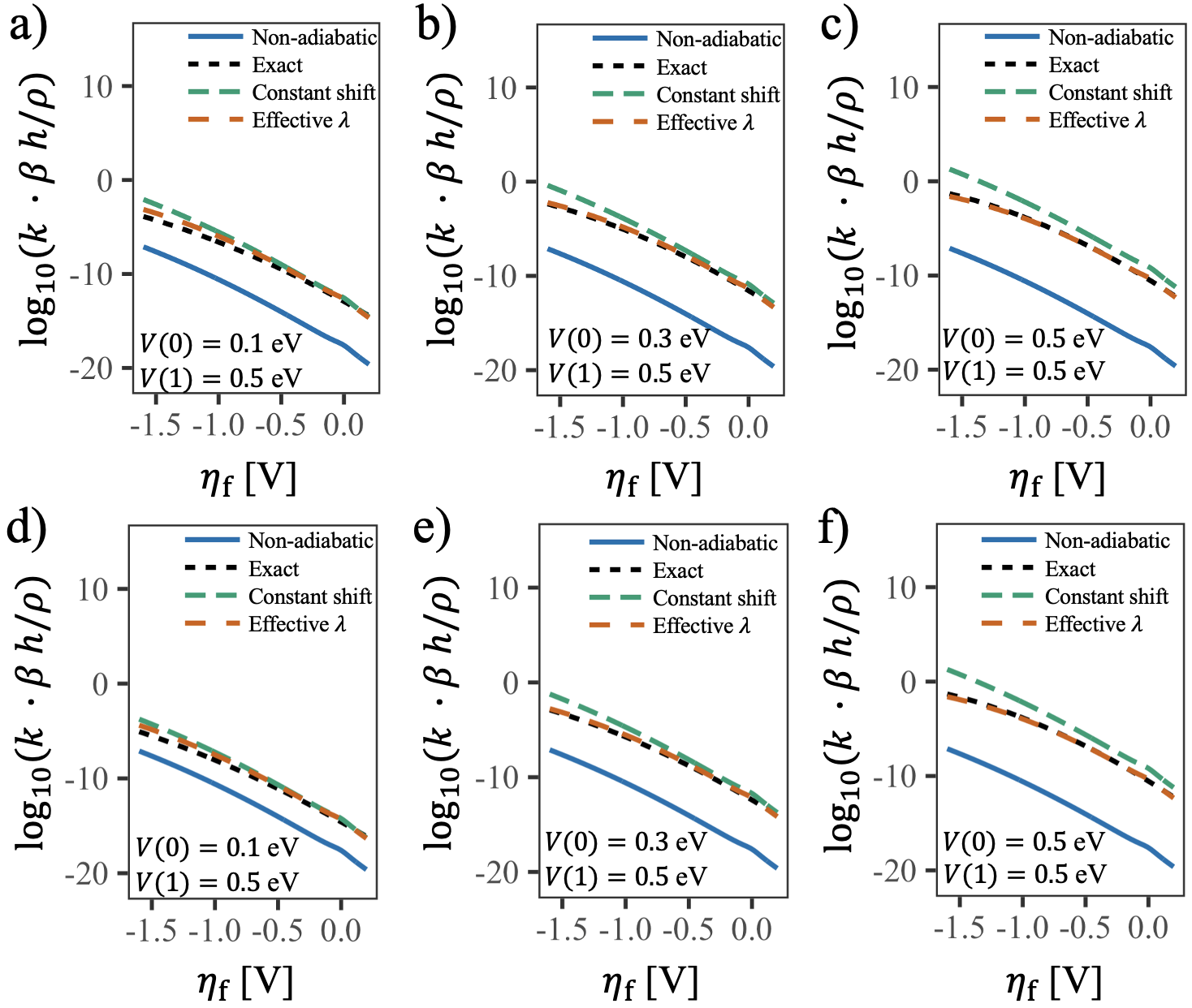}
\caption{Tafel plots showing the (base-10) logarithm of the (scaled) reaction rate $k$ plotted against the formal overpotential $\eta_\text{f}$ at $T=300$ K. Results are shown for (a)--(c) the linear $V(q)$ given by Eq. (\ref{linV}) and for (d)--(f) quadratic $V(q)$ given by Eq. (\ref{quadV}). Across each row of panels, the value $V(0)$ increased while the value of $V(1)$ is held fixed (note that plots (c) and (f) are equivalent). The orange curve is computed using Eq. (\ref{mhcapprox3})}
\label{FIGS3}
\end{figure}
 
 Finally, one may inquire how reasonable of an approximation it is to simply use $\lambda_\text{eff}(q=1/2)$ in Eq. (\ref{mhcapprox2}) and neglect also the dependence of $\lambda_\text{eff}$ on the overpotential. In this case, we obtain \begin{equation}\label{mhcapprox3}
 k=\frac{\rho\sqrt{\frac{\pi}{\beta}\left(\lambda-4V(0)+\frac{4\left[V(1/2)\right]^2}{\lambda}\right)}}{\beta h\left(1 + e^{\beta e \eta_\text{f}}\right)}
 \mathrm{erfc}\left(
\frac{\beta\left(\lambda-4V(0)+\frac{4\left[V(1/2)\right]^2}{\lambda}\right)- \sqrt{1 + \sqrt{\beta\left(\lambda-4V(0)+\frac{4\left[V(1/2)\right]^2}{\lambda}\right)} + (\beta e \eta_\text{f})^2}}
      {2\sqrt{\beta \left(\lambda-4V(0)+\frac{4\left[V(1/2)\right]^2}{\lambda}\right)}}
 \right).\end{equation} Figure S2 shows such results, again using the linear and quadratic forms for $V(q)$ in the top and bottom panels respectively. We see that when $V(0)$ is much less than $V(1),$ Eq. (\ref{mhcapprox3}) performs slightly worse at very large overpotentials, whereas as when $V(0)$ approaches $V(1)$, the accuracy is recovered. We see that the results for quadratic $V(q)$ perform slightly better than the linear case, and by inspection of its derivation, we can expect that Eq. (\ref{mhcapprox3}) will be a reasonable approximation as long as $(\partial V/\partial q)|_{q=1/2}$ is sufficiently small.

 \section{Adiabatic corrections to the thermodynamic driving force}

The careful reader may object that the thermodynamic driving force $\Delta E$ that enters into Eq. (\ref{ts}) is defined between the diabatic basins, and thus is not the same as the experimentally measured thermodynamic driving force $\Delta E_\text{eff}$ that should correspond to the difference between the adiabatic basins.

In the Condon approximation, it is simple to prove that $\Delta E_\text{eff}=\Delta E+O\left(V^k(\Delta E)^{3-k}/\lambda^2\right)$ for $k = 0,1,2,3$. This is easily proved by repeating the local expansion around the reactant minimum discussed above, now for the product minimum. We see that this is the same problem as the reactant basin upon making the substitutions $q\rightarrow 1-q,~\Delta E\rightarrow -\Delta E$. Thus, appropriately applying Eq. (\ref{eq:Er0_series_lambda_normal}) for both minima, we obtain \begin{equation}\label{eq:justify_geff}
\Delta E_\text{eff}=E^*(q_p)-E^*(q_\text{r})=\Delta E +O\left(\frac{V^2\,\Delta E}{\lambda^2},\frac{V^2(\Delta E)^2}{\lambda^3},\frac{V^4}{\lambda^3}\right),
\end{equation} 
which proves our desired result.

Outside the Condon approximation, differences may arise, and $\Delta E_\text{eff}\neq\Delta E.$ In particular, we will have \begin{equation}\label{delta_Geff}
\Delta E_\text{eff}-\Delta E\approx 4\left[\frac{V(0)^2}{\lambda}-\frac{V(1)^2}{\lambda}\right],\end{equation} and given that $\Delta E_\text{eff}$ is the relevant quantity for many measurements, calculations must be adjusted accordingly.

\end{document}